\documentclass[twocolumn,aps,prl,floatfix,superscriptaddress]{revtex4-1}
\usepackage{dingbat}
\usepackage{booktabs}
\usepackage{amsmath,amssymb}
\usepackage{graphicx,float}
\usepackage{times,txfonts}
\usepackage{color}
\usepackage{multirow}
\usepackage{bbold}
\usepackage{color}
\usepackage{soul}
\usepackage{hyperref}
\usepackage{times,txfonts}
\usepackage{natbib}
\usepackage{tikz}
\usetikzlibrary{backgrounds}

\newcommand{\bra}[1]{\left\langle #1 \right|}
\newcommand{\ket}[1]{\left| #1 \right\rangle}

\newcommand{\qo}[1]{``#1''}

\renewcommand{\epsilon}{\varepsilon}
\renewcommand{\phi}{\varphi}

\usepackage{sistyle}
\usepackage{setspace}
\usepackage{lipsum}
\usepackage{soul}
\definecolor{lightblue}{RGB}{185,210,248}
\sethlcolor{lightblue}

\begin{document}
\title{Weak Value Amplification Can Outperform Conventional Measurement in the Presence of Detector Saturation}
\author{J\'er\'emie Harris}
\email{jeremie.harris@uottawa.ca}
\affiliation{Max Planck Centre for Extreme and Quantum Photonics, Department of Physics, University of Ottawa, 25 Templeton St., Ottawa, Ontario, K1N 6N5 Canada}

\author{Robert W. Boyd}
\affiliation{Max Planck Centre for Extreme and Quantum Photonics, Department of Physics, University of Ottawa, 25 Templeton St., Ottawa, Ontario, K1N 6N5 Canada}
\affiliation{Institute of Optics, University of Rochester, Rochester, New York, 14627, USA}

\author{Jeff S. Lundeen}
\affiliation{Max Planck Centre for Extreme and Quantum Photonics, Department of Physics, University of Ottawa, 25 Templeton St., Ottawa, Ontario, K1N 6N5 Canada}

\begin{abstract}

Weak value amplification (WVA) is a technique in which one can magnify the apparent strength of a measurement signal. Some have claimed that WVA can outperform more conventional measurement schemes in parameter estimation. Nonetheless, a significant body of theoretical work has challenged this perspective, suggesting WVA to be \textit{fundamentally} sub-optimal. Optimal measurements may not be \textit{practical}, however. Two practical considerations that have been conjectured to afford a benefit to WVA over conventional measurement are certain types of noise and detector saturation. Here, we report a theoretical study of the role of saturation and pixel noise in WVA-based measurement, in which we carry out a Bayesian analysis of the Fisher information available using a saturable, pixelated, digitized, and/or noisy detector. We draw two conclusions: first, that saturation alone does \textit{not} confer an advantage to the WVA approach over conventional measurement, and second, that WVA \textit{can} outperform conventional measurement when saturation is combined with intrinsic pixel noise and/or digitization. 

\end{abstract}
\maketitle
Weak value amplification is an interference-based phenomenon originally proposed by Aharonov, Albert and Vaidman~\cite{aharonov:88}. It has been argued that WVA can amplify minute signals, thereby enabling the determination of small physical quantities that would otherwise be impractical to measure~\cite{hosten:08,dixon:09,loaiza:14,feizpour:11,turner:11,starling:09,pfeifer:11,hogan:11,zhou:12,starling:10,starling:10b,brunner:10,strubi:13,egan:12,zhu:11,hayat:13,steinberg:10}. The quantum metrological community is presently engaged in a contentious debate over the effectiveness of WVA as a parameter estimation technique~\cite{combes:15,knee:13,ritchie:91,pang:15,alves:15,pang:14,dressel:13,pang:14b}. As part of WVA, one designs a measurement that partitions a set of trials into a highly informative subset that occurs with low probability, and a less informative subset that occurs with high probability. The information-rich measurement subset is distinguished by an amplification of the signal of interest. Frequently, the less informative measurement subset is discarded outright, in a process known as post-selection. Although post-selection rejects only the least informative measurement outcomes, even these must carry \textit{some} (though potentially very little~\cite{viza:15}) useful information. The potential information loss that can accompany partitioning and post-selection has led many to suggest that WVA cannot be advantageous~\cite{combes:14,ferrie:14,tanaka:13,zhang:15} compared to optimal measurement strategies. This argument has been received by some with skepticism owing to the perceived success of WVA in many experiments~\cite{viza:15} and to a number of theoretical arguments that suggest that certain types of noise might make WVA advantageous in practice~\cite{jordan:14}. 

The case against WVA is rigorously cast in terms of Fisher information (FI), a quantity that provides a lower bound on the uncertainty associated with the measurement of a physical parameter~\cite{jaynes:03}. The FI can be determined from the probabilities of the various possible measurement outcomes of a particular experiment. In this sense, the Fisher information provides a measure of the quality of an experimental design; a well-designed experiment will elicit more FI about a parameter of interest than will a poorly-designed experiment. Related to FI is the concept of the quantum Fisher information (QFI), defined as the standard FI, maximized over the set of all possible measurement schemes~\cite{helstrom:69}. The QFI provides a theoretical upper bound on the standard FI, and can be achieved only in the limit of an optimal measurement. Thus far, many of the arguments criticizing WVA have appealed to the QFI~\cite{combes:14,tanaka:13,zhang:15}, which fails to account for the myriad practical constraints faced in real experiments. That the WVA approach fails to yield the full QFI \textit{theoretically} available to an experimenter should not, in and of itself, be understood to suggest WVA to be unhelpful \textit{in practice}. Rather than asking how WVA fares relative to the QFI, experimenters are concerned with the more practical question, \qo{Are any experimental constraints in my setup, that lead to an advantage for WVA schemes - and if so, what are they?} 

For example, one might imagine using a camera to measure a shift $g$ in the position of a laser beam's transverse spatial distribution, i.e. its \qo{profile}. We shall assume $g$ to be small compared to the beam width $w$, and that $w$ has been fixed by practical considerations. Via WVA, it is possible to amplify this shift considerably in exchange for post-selecting away a fraction of the photons in the beam, reducing its average photon number. A number of factors might conceivably make this trade-off worthwhile; some forms of noise, or the saturation of the camera's pixels might make the magnitude of the shift more important than the number of photons in the beam, for example. While study of beam jitter, detector jitter, turbulence and time-correlated noise has found WVA to be potentially advantageous~\cite{feizpour:11,jordan:14}, a rigorous treatment of saturation effects is absent from the literature. Detector saturation results in a flattening of the beam's spatial distribution, limiting one's ability to measure beam shifts. Intuitively, one might expect the WVA strategy to be more robust against such flattening of the beam's spatial distribution, since it simultaneously amplifies the shift and reduces the average photon number of the beam, in turn limiting detector saturation~\cite{vaidman:14}. Surprisingly, however, this argument does not hold as stated.

Here, we present a rigorous treatment of the case of detector saturation. We analyze saturation, digitization, intrinsic pixel noise and/or pixelation in a camera, which is used to measure the profile of a laser beam. That is, a beam whose quantum optical state is a coherent state with average photon number $\bar{n}$. From this example, we draw two surprising and general conclusions: first, that saturation alone does \textit{not} confer an advantage to WVA, but second, that saturation, in concert with intrinsic pixel noise and/or digitization, \textit{can} make WVA advantageous relative to a conventional measurement. Our findings clarify any perceived disagreement between numerous theoretical works that purport to show WVA to be suboptimal, and the perspective of many theorists and experimentalists, to whom the WVA technique has proven itself in principle and in practice to be a superior alternative to conventional measurement schemes in many situations.

We begin by considering the \qo{conventional} measurement (CM) of $g$. For concreteness, we take $g$ to represent the transverse displacement of a beam's spatial distribution, brought about by its having propagated through a birefringent material. The material's birefringence causes the beam's polarization to become correlated with its position, so that the beam undergoes a transformation given by $\ket{i}\ket{\psi} \rightarrow e^{-ig\hat{A}\hat{P}_x}\ket{i}\ket{\psi}$, where we denote by $\ket{i}$ and $\ket{\psi}$ the initial polarization and transverse distribution of the beam, and $\hat{A}$ and $\hat{P}_x$ respectively represent the operators associated with the beam's polarization and transverse momentum in the $x$-dimension. The states $\ket{i}$ and $\ket{\psi}$ describe the optical modes occupied by the coherent state. For CM, one might choose $\ket{i}=\ket{H}$, meaning that the beam is prepared in a horizontal polarization state, which is taken to be an eigenstate of $\hat{A}$ with eigenvalue $+1$. After propagating through the birefringent material, the beam's state then becomes $e^{-ig\hat{A}\hat{P}_x}\ket{H}\ket{\psi} = e^{-ig \hat{P}_x} |H\rangle |\psi\rangle=|H\rangle|\psi_\text{CM}\rangle$. Given that $e^{-ig \hat{P}_x}$ is the spatial translation operator, the photon's spatial profile is shifted by $g$. More explicitly, the photon's \qo{spatial amplitude} in the CM case will be $\langle x |\psi_\text{CM}\rangle = \psi(x-g)$. $g$ can therefore be measured by observing the beam's spatial profile with a camera, i.e. $\bar{n}|\psi(x-g)|^2$.

%
\begin{figure*}[t]
	\includegraphics[width=2\columnwidth]{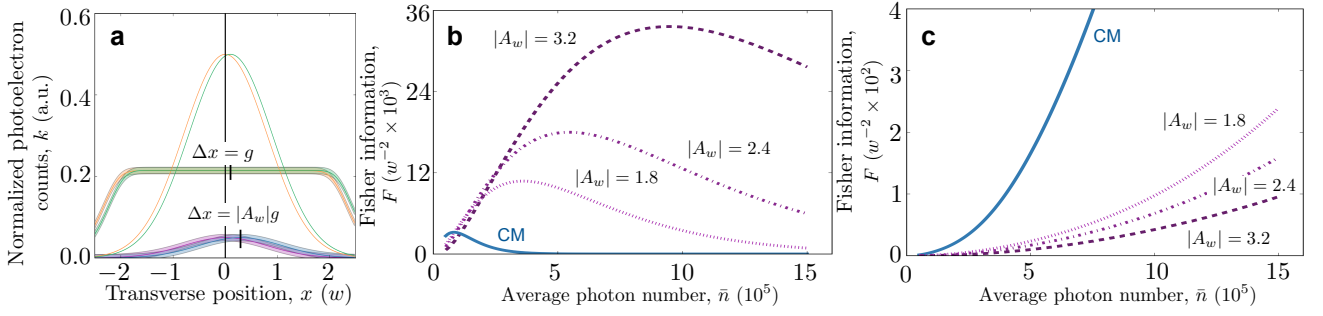}
	\caption{{\bf Saturation is necessary to afford an advantage to weak value amplification over conventional measurement.} a) Beam distortion produced by saturation for both CM and WVA with $g|A_w|=0.364$, for a beam with waist $w=0.8$. Orange and green lines show the beam profiles impinging on the camera, while the corresponding beam measured beam profiles obtained from a noisy and saturable camera are indicated by the shaded orange and green profiles. Measured beam profiles obtained via WVA (shaded magenta and blue) are dimmer, though more robust against saturation effects. b) FI obtained from a camera with $k_\text{max}=256$, $\sigma=k_\text{max}/20$ and $N_\text{sat}=500$, for an incident beam with $w=1$ and $g=0.01$, carrying a range of average photon numbers. FIs obtained for CM (solid blue) are plotted alongside FIs for indicated WVA schemes (dotted/dashed purple). The plot  shows that more FI can be recovered using WVA than using CM when saturation effects become important, meaning that WVA can outperform CM in this regime.  c) FI obtained for a beam and camera identical to that simulated in b), except that $N_\text{sat}=10^5$, in order to render saturation negligible. Without saturation, but with intrinsic pixel noise, pixelation and digitization, WVA is outperformed by CM for all $\bar{n}$ plotted.} 
	\label{fig:1}
\end{figure*}

The WVA scheme differs from CM in two respects. First, it requires that the laser beam's polarization be prepared in in a superposition of orthogonal polarization states. We consider the case $\ket{i}=\frac{1}{\sqrt{2}} (\ket{H}+\ket{V})$, where $\ket{V}$ denotes vertical polarization. Second, it involves post-selecting on the beam's polarization once it emerges from the birefringent material, so that only outcomes for which the polarization matches some particular state $\ket{f}$ are retained. The (unnormalized) position state becomes $\ket{\psi_\text{WVA}} \approx \bra{f} e^{-ig\hat{A}\hat{P}_x}\ket{i}\ket{\psi}$. For small $g$, $\ket{\psi_\text{WVA}} \approx \bra{f}(1 -ig\hat{A}\hat{P}_x )\ket{i}\ket{\psi} \approx \langle f | i \rangle \, e^{- ig A_w \hat{P}_x} |\psi\rangle$. Again, the beam is spatially shifted, but now by $A_w g$, where $A_w = \langle{f} | \hat{A} | {i} \rangle / \langle{f} | i \rangle$ is the weak value of $\hat{A}$, taken to be purely real by choosing an appropriate $\ket{i}$ and $\ket{f}$. $A_w$ can, in principle, take on values well outside of the eigenspectrum of $\hat{A}$. In WVA, one chooses $\ket{i}$ and $\ket{f}$ to be nearly orthogonal, so that the denominator in the weak value expression becomes small, leading to a large $A_w$ and a magnified  shift $A_w g$. Physically, this arises from the interference between pointer states correlated with the two polarizations $|H\rangle$ and $|V\rangle$. The resulting beam spatial profile is then $\bar{n} |\langle x | \psi_\text{WVA} \rangle|^2 = \bar{n}p_\text{ps} |\psi(x-g A_w)|^2$, where $p_\text{ps} = |\langle f | i \rangle|^2$ is the post-selection probability, and $\psi(x)$ is spatially normalized. This $A_w$-fold increase in the shift of the beam's transverse distribution is precisely the signal magnification conjectured to offer WVA an advantage over CM. Despite this amplification effect, however, the average photon number reaching the camera is reduced by a factor $p_\text{ps}$. This photon loss competes with the signal amplification effect to determine the accuracy afforded by the WVA technique overall (see Figure \ref{fig:1}a). 

We now consider an experiment in which a camera is used to measure the small beam shift $g$, though our treatment can be generalized to cases in which any two degrees of freedom possessed by a beam are coupled. Our experiment will admit a set $\{k\}$ of possible outcomes, which occur with probability $p(k|g,X)$. Here, $X$ represents all available prior information, including the saturation model under consideration (if any), as well as camera pixelation, digitization and intrinsic pixel noise effects. The FI associated with this experiment is ~\cite{jaynes:03}

%
\begin{align}\label{eq:FisherInfo}
\begin{aligned}
	F(g,X) = \sum_k p(k|g,X) \left( \frac{\partial}{\partial g} \ln{p(k|g,X)} \right)^2.
\end{aligned}
\end{align}

\noindent A photon incident on the camera can produce a range of photoelectron counts at the pixel at which it \qo{arrives}. Hence, a particular experimental outcome $k=\{k_1,k_2,...,k_M\}$ refers to a number of photoelectrons excited at each of the camera's $M$ pixels, where we define $k_j$ as the proposition, \qo{$k_j$ photoelectrons were excited at pixel $j$.} 

For a beam in a coherent state, the distribution of photon numbers reaching each pixel is independent and Poissonian~\cite{barnett:97}. The distribution of photon numbers reaching pixel $j$, $N_j$, is therefore 

%
\begin{align}\label{eq:PhotonProb}
\begin{aligned}
	p(N_j|g,X) = \frac{\bar{n}_j^{N_j} \, e^{-\bar{n}_j}}{N_j !},
\end{aligned}
\end{align}

\noindent where we define the average photon number reaching pixel $j$ as $\bar{n}_j=\bar{n}\int_j |\langle x | \psi_\text{CM}\rangle|^2 \text{d}x$ for CM, or $\bar{n}\int_j |\langle x | \psi_\text{WVA}\rangle|^2 \text{d}x$ for WVA. This approach naturally accounts for photon losses that result from the finite camera size. The mutual independence of each pixel allows the Fisher information at each pixel, $F_j(g,X)$, to be determined independently, such that the total FI available to the camera becomes $F(g,X) = \sum_{j=1}^M F_j(g,X)$~\cite{jaynes:03}. To find the FI available from the camera, we thus need concern ourselves only with $p(k_j|g,X)$. 

The probability of exciting $k_j$ photoelectrons at camera pixel $j$ can be determined from the probabilities that $N_j$ photons from the beam will reach that pixel, as follows:

%
\begin{align}\label{eq:BayesProb}
\begin{aligned}
	p(k_j|g,X) = \sum_{N_j} p(k_j|N_j,g,X) \, p(N_j|g,X).
\end{aligned}
\end{align}

\noindent We assume our prior information $X$ to include the average photon number $\bar{n}$ of the initial optical beam. 

The quantity $p(k_j|N_j,g,X)$ is the probability of exciting $k_j$ photoelectrons given that $N_j$ photons impinge on pixel $j$, and accounts for saturation effects, pixel digitization and intrinsic pixel noise in our treatment. We introduce intrinsic noise at each pixel by modelling the photoelectron response $k_j  \sim \mathcal{N}(\mu_j,\sigma^2)$, as a normal distribution with mean $\mu_j$ and variance $\sigma^2$. We note that pixel noise should be distinguished from detector transverse jitter, angular beam jitter and turbulence, which are treated in~\cite{jordan:14}, and which affect the beam itself, rather than the detector. Hence, 

%
\begin{align}\label{eq:PhotoelProb}
\begin{aligned}
	p(k_j|N_j,g,X) = \frac{e^{-(k_j - \mu_j)^2/2\sigma^2}}{\sum_{k'_j} e^{-(k'_j - \mu_j)^2/2\sigma^2}}.
\end{aligned}
\end{align}

\noindent We include saturation effects by choosing $\mu_j = k_\text{max} \left( 1 - e^{-N_j/N_\text{sat} } \right)$, so that the photoelectron response scales linearly with $N_j$ for  beams with low $\bar{n}$, saturating once $N_j \gtrsim N_\text{sat}$. The probabilities $p(k_j|g,X)$ are evaluated by combining the results (\ref{eq:PhotonProb}) and (\ref{eq:PhotoelProb}) via (\ref{eq:BayesProb}), from which the FI is obtained. We illustrate combined effects of detector noise and saturation in Figure \ref{fig:1}a, where $\sigma$ results in uncertainty regarding the magnitude of the beam's spatial profile at each point along the beam cross-section. 

A range of $N_j$ values is mapped probabilistically to a particular $k_j$. Thus, $k_\text{max}$ indicates the number of resolvable ranges of $N_j$ accommodated by a particular camera, providing a measure of the digitization of each pixel; a larger $k_\text{max}$ describes a camera with more bits. The intrinsic pixel noise is modelled by $\sigma$, and the effects of pixelation are controlled by changing $M$. Finally, each pixel's sensitivity to saturation can be tuned via $N_\text{sat}$.It thus becomes possible to investigate the effects of camera pixelation, intrinsic pixel noise, digitization and saturation on the Fisher information available using CM and WVA. 

%
\begin{figure}[t]
	\centering
	\includegraphics[width=\columnwidth]{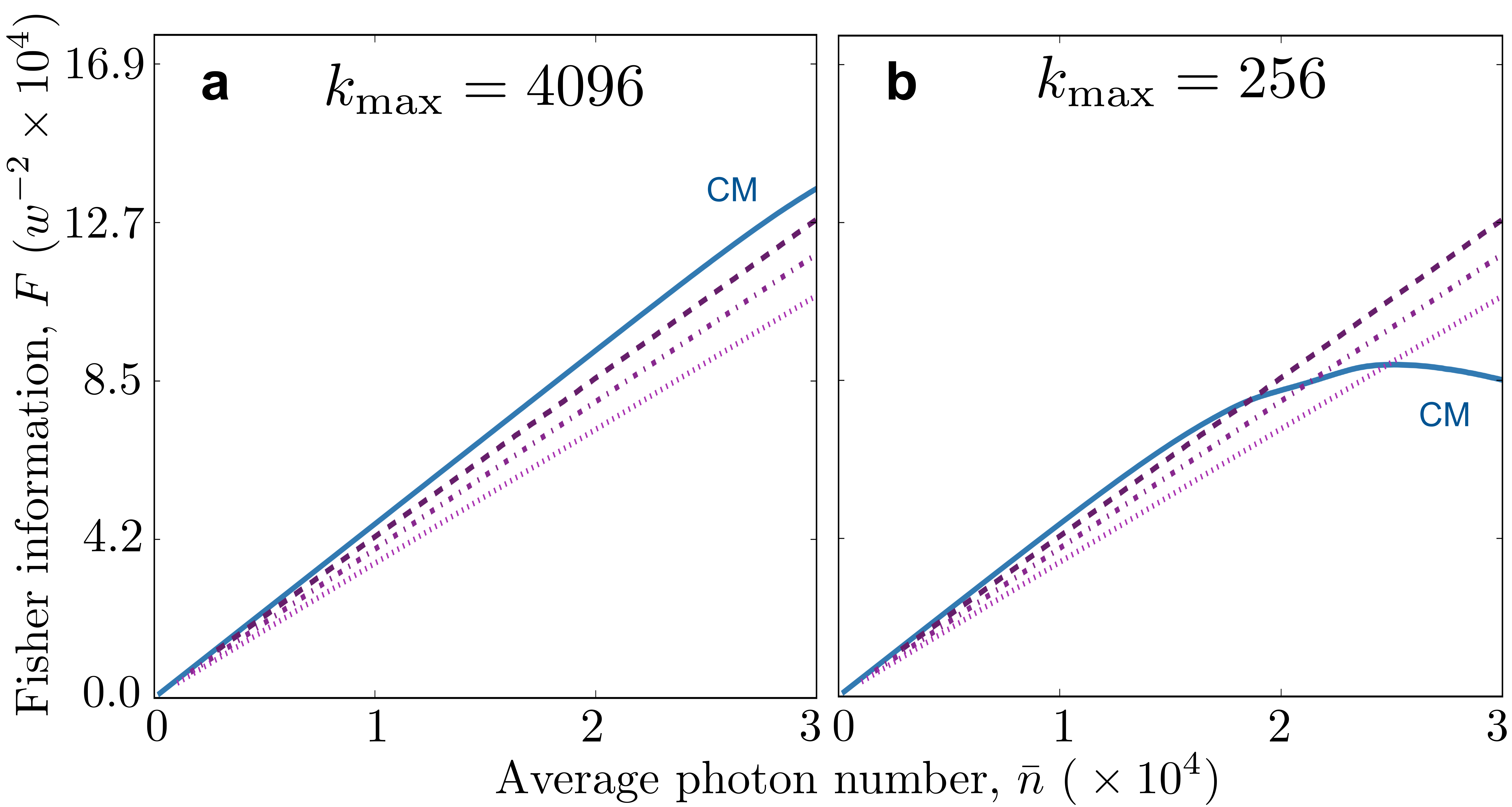}
	\caption{{\bf Saturation is not sufficient to afford an advantage to weak value amplification.} a) FI for a saturable ($N_\text{sat}=500$), noiseless, 100-pixel camera subject to negligible digitization ($k_\text{max}=4096$), for a beam identical to that simulated in Figure \ref{fig:1}c, for a range of $\bar{n}$ values. Despite the presence of saturation, CM measurement (blue) outperforms WVA with $|A_w|=1.8$ (dotted purple), 2.4 (dotted-dashed purple) and 3.2 (dashed purple) for all $\bar{n}$. b) Simulation identical to a), with the exception that digitization has been introduced by setting $k_\text{max}=256$. Although saturation alone fails to confer an advantage to WVA over CM (part a)), WVA can outperform CM when digitization is introduced (part b)).} 
	\label{fig:2}
\end{figure}
\begin{figure}[t]
	\centering
	\includegraphics[width=\columnwidth]{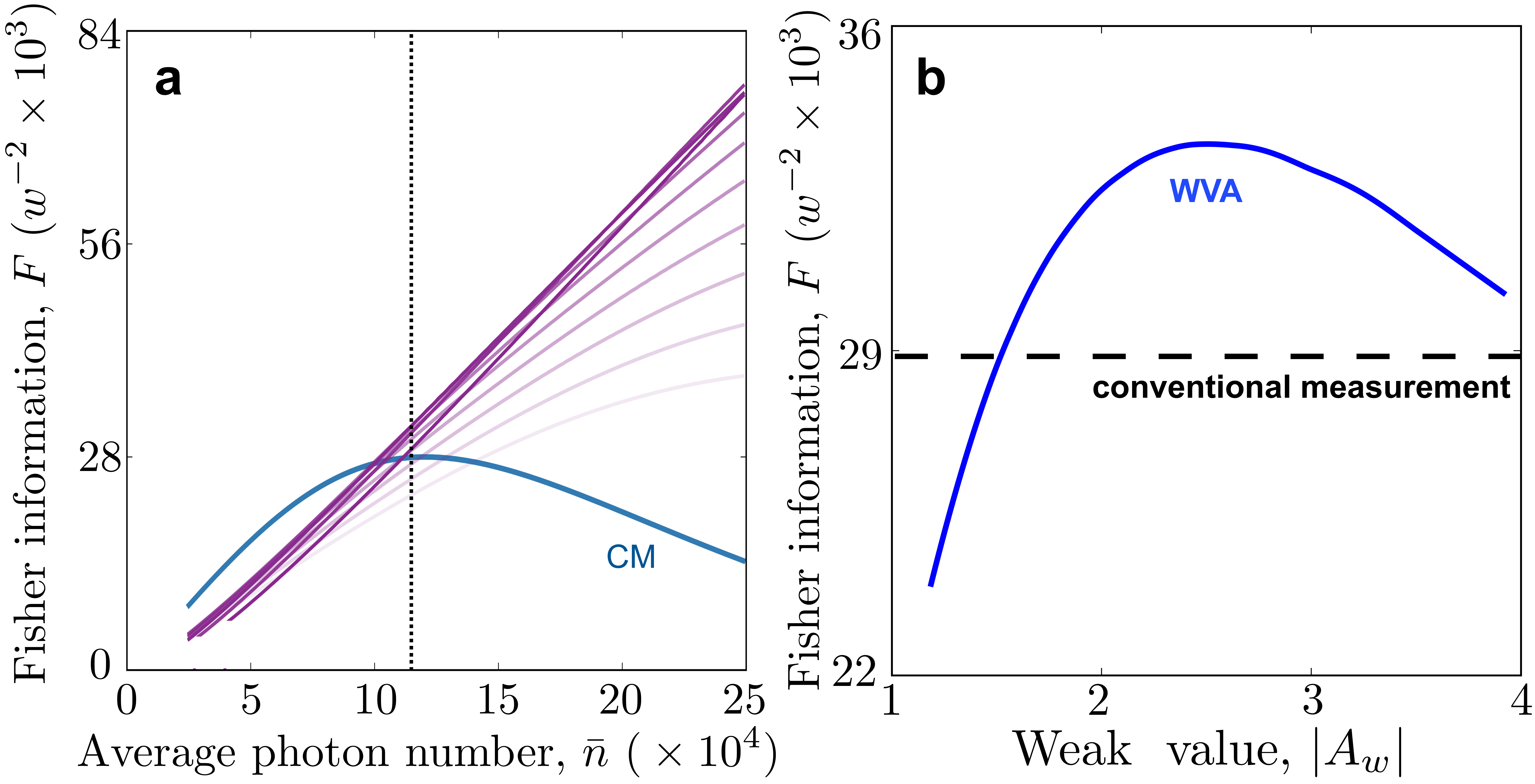}
	\caption{{\bf Optimal post-selection with weak value amplification.} a) Performance of CM (blue) and WVA schemes with amplification factors of $1.2$ (lightest purple), $1.3$, $1.5$, $1.6$, $1.8$, $2.1$, $2.4$, $2.7$, $3.2$ and $3.9$ (darkest purple) for a 100-pixel camera with $N_\text{sat}=500$, $k_\text{max}=256$, and $\sigma=k_\text{max}/100$ for indicated incident beam brightnesses. b) FIs for the camera simulated in a), for an incident beam with $\bar{n}=11500$ (dotted line in a)), showing that an optimal amplification $|A_w|$ exists, which elicits the maximum possible FI from the camera. The dashed line indicates the performance of CM.} 
	\label{fig:3}
\end{figure}

We now consider a realistic saturable camera, further limited by digitization and intrinsic pixel noise. We choose $k_\text{max} = 256$ (an 8-bit camera), a saturation number $N_\text{sat} = 500$, a noise parameter $\sigma = k_\text{max}/20$, and a Gaussian beam $|\psi(x)|^2 = e^{-x^2/2w^2}/2\pi w$, with $w = 1$. In Figure \ref{fig:1}b, we plot the total FI available to this camera, for incident beams with various $\bar{n}$ values, using CM and WVA with $|A_w| = 1.8, 2.4$ and $3.2$. For low $\bar{n}$, saturation plays only a minor role in determining the per-pixel FI, and CM outperforms each of the WVA strategies under consideration. As $\bar{n}$ increases, however, saturation effects begin to compromise the performance of CM, while WVA, which is more robust against saturation, becomes increasingly viable. Once $p(j|g) \, \bar{n}$ begins to considerably exceed $N_\text{sat}$, increases in $\bar{n}$ hinder one's ability to measure the transverse shift of the beam's spatial distribution, as the beam's image is prohibitively deformed. In this regime, increases in $\bar{n}$ actually \textit{reduce} the FI available to the camera. The low FIs associated with the high-$\bar{n}$ regime arise from a \qo{washing out} of the beam profile observed using the camera. As $\bar{n} \rightarrow \infty$, the camera records a perfectly flat beam profile that carries no information about $g$ whatsoever. We note that even the WVA schemes considered in the figure are eventually subject to saturation, as $p_\text{ps} \, p(j|g) \, \bar{n}$ overtakes $N_\text{sat}$, but the maximum FI retrievable from WVA, $F_\text{WVA}(\bar{n}_\text{WVA}^\text{max})$, always exceeds that from CM, $F_\text{CM}(\bar{n}_\text{CM}^\text{max})$. When photons are not in short supply, therefore, one invariably benefits from WVA. This is no longer true when the effects of saturation are removed, as in Figure \ref{fig:1}c. Hence, saturation is \textit{necessary} for WVA to outperform CM; one may be tempted to also assume that saturation is \textit{sufficient}. However, we will now show analytically that this is not the case. 

In the absence of digitization or intrinsic pixel noise, a pixel's photoelectron count $k_j$ is associated with a unique number of incident photons $N_j$. Hence, $p(k_j|g,X)=p(N_j|g,X)$. Since the probabilities of experimental outcomes are unaffected by saturation, saturation alone cannot affect the FI of a pixelated camera. As discussed in Ref~\cite{knee:14}, pixelation alone cannot lead to an advantage for WVA over schemes that make no use of post-selection. Saturation and/or pixelation are therefore insufficient to allow WVA to outperform CM. Rather, the superior performance of WVA in the high $\bar{n}$ regime of Figure \ref{fig:1}c arises from the combination of saturation, intrinsic pixel noise and digitization. 

Figure \ref{fig:2}a confirms that a noiseless camera in the near-absence of digitization ($k_\text{max}=4096$) will not confer an advantage to WVA over CM, even in the presence of saturation, for various $\bar{n}$. However, when digitization is introduced by setting $k_\text{max}=256$, the combined presence of saturation and digitization can make WVA advantageous (Figure \ref{fig:2}b). We note that digitization and intrinsic pixel noise play similar roles in our model, as both reduce the information available about the parameter $g$, by eliminating the one-to-one mapping between incident photon number at a pixel, $N_j$, and the number of photoelectrons excited at the pixel, $k_j$, that otherwise guarantees the superior performance of CM. Our results therefore indicate that saturation alone fails to confer an advantage to WVA over CM, but that when this effect is paired with digitization or intrinsic pixel noise, as is the case in virtually any experiment, WVA can outperform CM considerably. These conclusions are summarized in Table \ref{table:1}.

Finally, in Figure \ref{fig:3}, we show that an optimal WVA amplification factor $|A_w|$ exists for a particular beam brightness, provided that the camera saturates. When $|A_w|$ is small, increases in the amplification factor lead to increases in FI, as post-selection amplifies the shift of the beam's spatial distribution, and reduces beam distortion from saturation. However, beyond a certain optimal $|A_w|$ value, the measured beam is prohibitively dimmed, resulting in a net loss of information. 

One might suggest that the poor performance of CM in the high-saturation regime can be addressed by dimming the laser source, or introducing an attenuator between the source and camera. Even then, however, WVA will always outperform CM in this regime since it makes use of a post-selection procedure that retains the maximally informative photons from the beam. If one \textit{must} discard photons, the WVA method will always produce a result superior to that obtained from indiscriminate post-selection. 

\begin{table}[h]
\centering
$\begin{array}{ *{8}{c} }
\toprule
 & Saturation & Digitization & Pixel\, noise & Pixelation \\
\midrule
Saturation & $X$ & $\checkmark$ & $\checkmark$ & $X$   \\ 
Digitization & $\checkmark$ & $X$ & $X$ & $X$  \\ 
Pixel\, noise & $\checkmark$ & $X$ & $X$ & $X$  \\ 
Pixelation & $X$ & $X$ & $X$ & $X$  \\ 
\bottomrule
\end{array}$
\caption{{\bf Conditions under which weak value amplification can outperform conventional measurement.} Experimental circumstances under which WVA can potentially outperform CM are indicated by a \checkmark, whereas situations in which WVA cannot outperform CM are indicated by a X. Pixelation includes results from ref~\cite{knee:14}.} 
\label{table:1}
\end{table}

We have shown that weak value amplification can outperform conventional measurement when intrinsic pixel noise and digitization are introduced with our saturation model, although each of these effects independently fails to provide an advantage for the weak value technique. We note, that while the Fisher information does provide an upper bound on the precision of parameter estimation in a particular experiment, this bound can be reached only using the maximum likelihood estimator (MLE). The MLE may be impractical to implement, requiring detailed knowledge of the dimensions, noise, saturation and digitization characteristics of one's camera. Nonetheless, we speculate that a straightforward center-of-mass estimator~\cite{vaughan:07} will approach the sensitivity of the MLE. This would require sufficiently large weak value amplification to ensure that the signal is attenuated enough so that camera's response is linear. In addition, biasing factors such as noise and clipping of the beam by the camera edges will need to be avoided or accounted for.

We have also found that, for sufficient beam brightnesses, a unique amplification factor $|A_w|$ maximizes the Fisher information retrievable from a realistic camera. Apart from their practical significance, our results are of fundamental interest, as they provide conclusive and surprising answers to questions surrounding the possible benefits of weak value amplification in the presence of saturation. They also offer a lens through which many apparently contradictory claims made in the literature about the potential benefits of weak value amplification can finally be reconciled.

\begin{acknowledgments} J.H., J.L. and R.W.B. acknowledge the support of the Canada Excellence Research Chairs (CERC) Program and NSERC. J.H. acknowledges the support of the Vanier Canada Graduate Scholarships Program. J.L. acknowledges the support of the Canada Research Chairs (CRC) Program. We would like to thank Edouard Harris for his assistance designing the FI algorithm, and A. Jordan for helpful comments on the manuscript. \end{acknowledgments}

\end{document}